\documentclass[prl,twocolumn]{revtex4}
\usepackage{graphicx}
\usepackage{amssymb,amsmath,bm}
\usepackage{enumerate}

\newcommand{\be}{\begin{eqnarray}}
\newcommand{\ee}{\end{eqnarray}}

\begin{document}

\title{Distributed chaos tuned to large scale coherent motions in turbulence}

\author{A. Bershadskii}

\affiliation{
ICAR, P.O. Box 31155, Jerusalem 91000, Israel
}

\begin{abstract}

It is shown, using direct numerical simulations and laboratory experiments data, that distributed chaos is often tuned to large scale coherent motions in anisotropic inhomogeneous turbulence. The examples considered are: fully developed turbulent boundary layer (range of coherence: $14 < y^{+}  < 80$), turbulent thermal convection (in a horizontal cylinder), and Cuette-Taylor flow. Two ways of the tuning have been described: one via fundamental frequency (wavenumber) and another via subharmonic (period doubling). For the second way the large scale coherent motions are a natural component of distributed chaos. In all considered cases spontaneous breaking of space translational symmetry is accompanied by reflexional symmetry breaking.
\end{abstract}

\maketitle

\section{Introduction}

Large scale coherent motions (LSCM) and their interaction with comparatively small scale turbulence in anisotropic inhomogeneous flows is subject of many studies and speculations. Let us begin our investigation from a short consideration of analogous problem in deterministic chaos. We will see that the main features observed in this case are analogous to those in turbulence driven by distributed chaos. 
\begin{figure}
\begin{center}
\includegraphics[width=8cm \vspace{-1cm}]{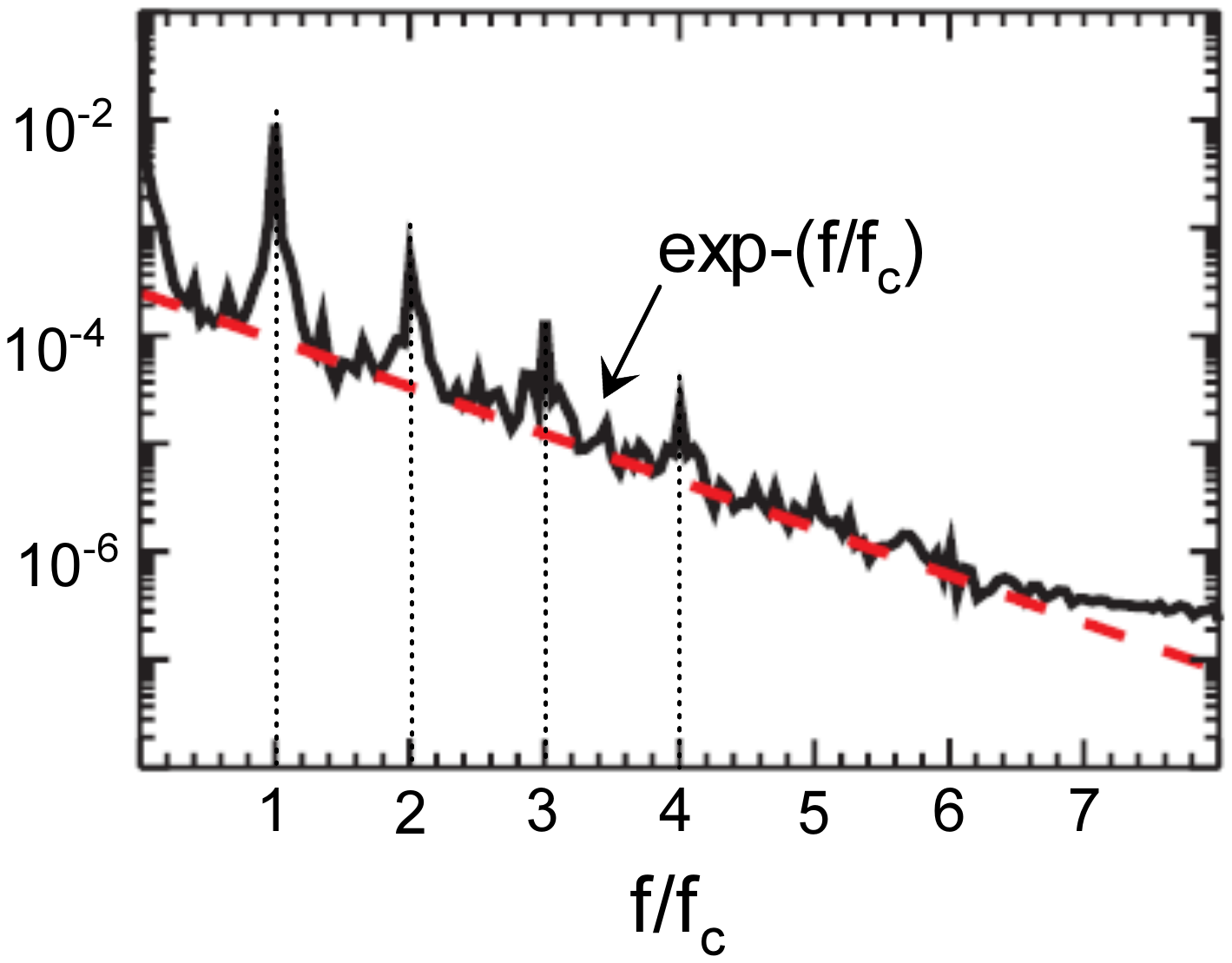}\vspace{-5cm}
\caption{\label{fig1} Logarithm of the temperature fluctuations power spectrum against normalized frequency $f/f_c$ (adapted from the Reference \cite{mm}). The dashed straight line corresponds to Eq. (1) with $f_{\beta}=f_c$ ($f_c$ is the fundamental frequency of the drift waves).  }
\end{center}
\end{figure}
\begin{figure}
\begin{center}
\includegraphics[width=8cm \vspace{-1cm}]{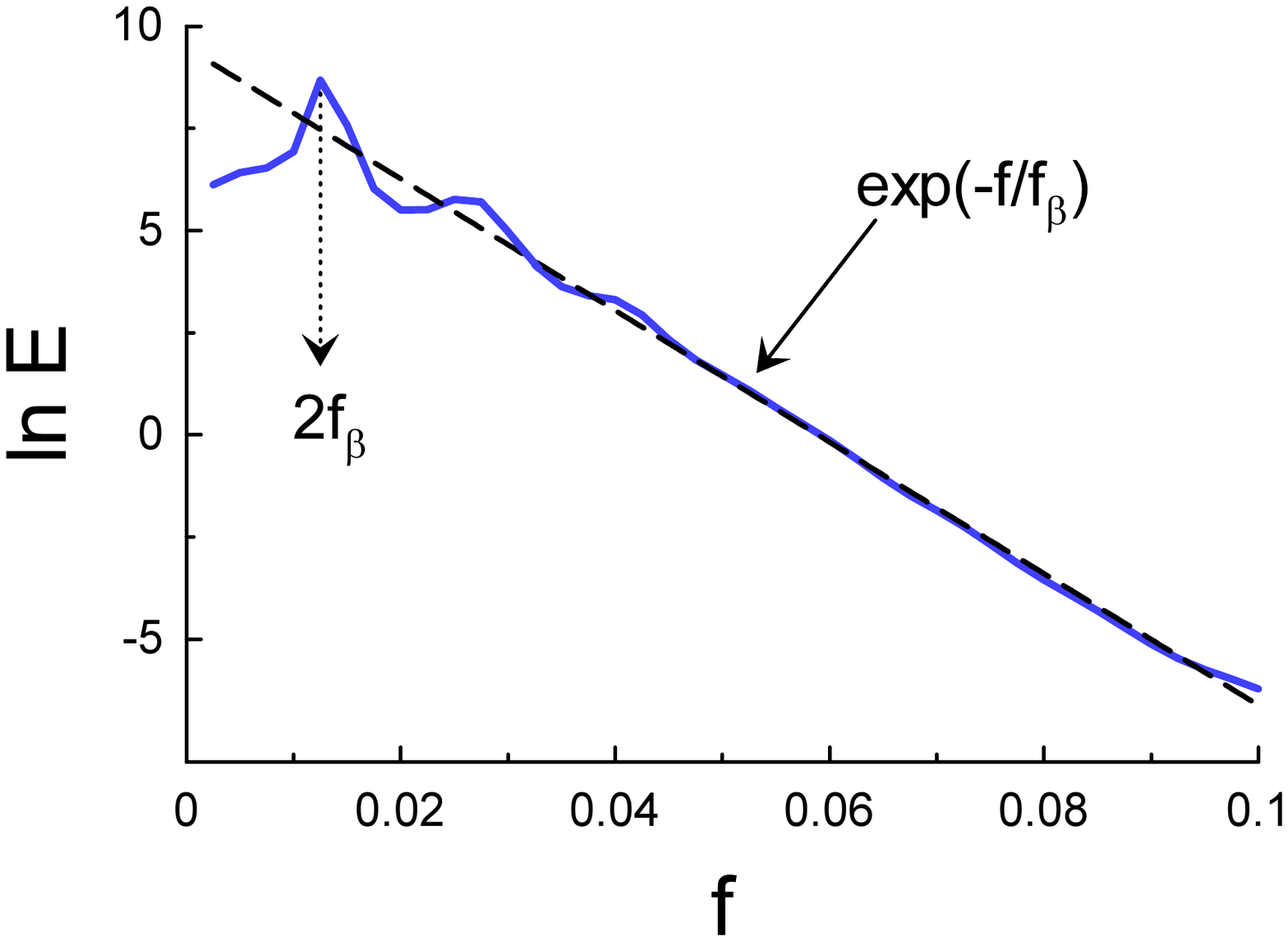}\vspace{-4.3cm}
\caption{\label{fig2} Logarithm of power spectrum for $z$-component of the Eq. (2) against frequency $f$. The dashed straight line corresponds to Eq. (1).} 
\end{center}
\end{figure}
  The smooth deterministic chaos is often characterized by exponentially decaying broadband spectrum \cite{fm}-\cite{mm}
$$
E(f) \propto e^{-f/f_{\beta}}  \eqno{(1)}
$$    
  The most clear manifestation of the interaction between the LSCM and small scale chaotic motions is a tuning of the exponential decay parameter to the main manifestation of the LSCM - a low-frequency peak.  
    
    A two-mode simulation \cite{mm} of a temperature filament relaxation in magnetized plasma (after onset of chaos) provides a good example of such tuning. In this simulation unstable drift waves play role of the LSCM, which drive the chaos. In Figure 1 one can clear see the tuning of the exponential decay Eq. (1) to the fundamental frequency of the drift waves $f_c = f_{\beta}$ (for more information and examples see Ref. \cite{b1}). \\

  The seminal Lorenz equations: 
$$
\frac{dx}{dt} = \sigma (y - x),~~      
\frac{dy}{dt} = r x - y - x z, ~~
\frac{dz}{dt} = x y - b z      \eqno{(2)}          
$$
provide another example of the tuning. The Eqs. (2) correspond to a simplified model for thermal (Rayleigh-Benard) convection in a layer of fluid, heated from below and cooled from above \cite{lorenz}. Usually used values of the parameters $\sigma=10.0,~ r = 28.0,~ b = 8/3$ result in a chaotic solution \cite{eck}. In Fig. 2 we show a power spectrum of $z$-component in Eq. (2). The spectrum was computed by the maximum entropy method. This method gives an optimal resolution for comparatively short time series (in this case $3\cdot 10^4$ data points). Unlike the previous example the tuning here is subharmonic (period-doubling): the fundamental frequency, corresponding to the main spectral peak, is equal to $2f_{\beta}$. The LSCM, generating the main spectral peak, in this case are a natural part of the deterministic chaos.

\section{Distributed chaos in turbulence}

  In distributed chaos the value of $f_{\beta}$ in Eq. (1), instead being tuned, is broadly distributed. Corresponding spectrum is a weighted superposition of the exponentials
$$
E(f) \propto \int P(f') e^{-f/f'} df'  \eqno{(3)}
$$  
where $P(f')$ is a probability distribution. It is shown in Ref. \cite{b2} that in turbulence the integral in the right-hand side of Eq. (3) is converged into a stretch exponential
$$
E(f) \propto \int P(f') e^{-f/f'} df' \propto \exp -(f/f_{\beta})^{\beta} \eqno{(4)}
$$ 
In this case the $f_{\beta}$ from the stretched exponential Eq. (4) should be considered for a tuning to the LSCM.

   For turbulence the Taylor hypothesis \cite{my} states that the frequency spectra can be transformed (at certain conditions) into wave-number spectra ($f \rightarrow k$):
$$
E(k) \propto \int P(\kappa) e^{-k/\kappa} d\kappa \propto \exp -(k/k_{\beta})^{\beta} \eqno{(5)}
$$ 
and the tuning should be considered for the value of $k_{\beta}$.

   For homogeneous isotropic turbulence momentum correlation integral 
$$
\int  \langle {\bf u} ({\bf x},t) \cdot  {\bf u} ({\bf x} + {\bf r},t) \rangle d{\bf r}~ \longrightarrow~ \beta = 3/4  \eqno{(6)}
$$
 (Birkhoff-Saffman invariant \cite{saf}\cite{d}) controls distributed chaos and $\beta = 3/4$ in Eqs. (4) and (5) \cite{b2}. This invariant is a consequence of space translational symmetry - homogeneity (according to Noether's theorem \cite{ll}), and spontaneous breaking of this symmetry switches control to vorticity correlation integral 
$$
\int_{V} \langle {\boldsymbol \omega} ({\bf x},t) \cdot  {\boldsymbol \omega} ({\bf x} + {\bf r},t) \rangle_{V}  d{\bf r} ~\longrightarrow ~\beta = 1/2  \eqno{(7)}
$$
with $\beta = 1/2$ in the case of vortex like LSCM domination \cite{b3}. In the case of (helical) wave like LSCM domination the control is switched to helicity correlation integral 
$$
\int_V \langle h({\bf x},t)~h({\bf x}+{\bf r}, t)\rangle_V d{\bf r}~\longrightarrow ~\beta = 1/3 \eqno{(8)}
$$
(Levich-Tsinober invariant \cite{lt},\cite{fl},\cite{l}) with $\beta=1/3$ \cite{b4}. 
   
\section{Fully developed turbulent boundary layer}

Usually (and rather vaguely) the near-wall region of fully developed turbulent boundary layer has been divided into three main sublayers: the laminar (viscous) sublayer, the buffer layer ($10 < y^+ < 40$) and the logarithmic layer (where $y^{+}= yu_{\tau}/\nu$ is dimensionless distance to the wall, with the friction velocity $u_{\tau}$ and kinematic viscosity $\nu$). It is believed that the main turbulence production occurs in the buffer region, where the most near-wall coherent structures are located \cite{jeo},\cite{smits},\cite{dennis}.
\begin{figure}
\begin{center}
\includegraphics[width=8cm \vspace{-1.3cm}]{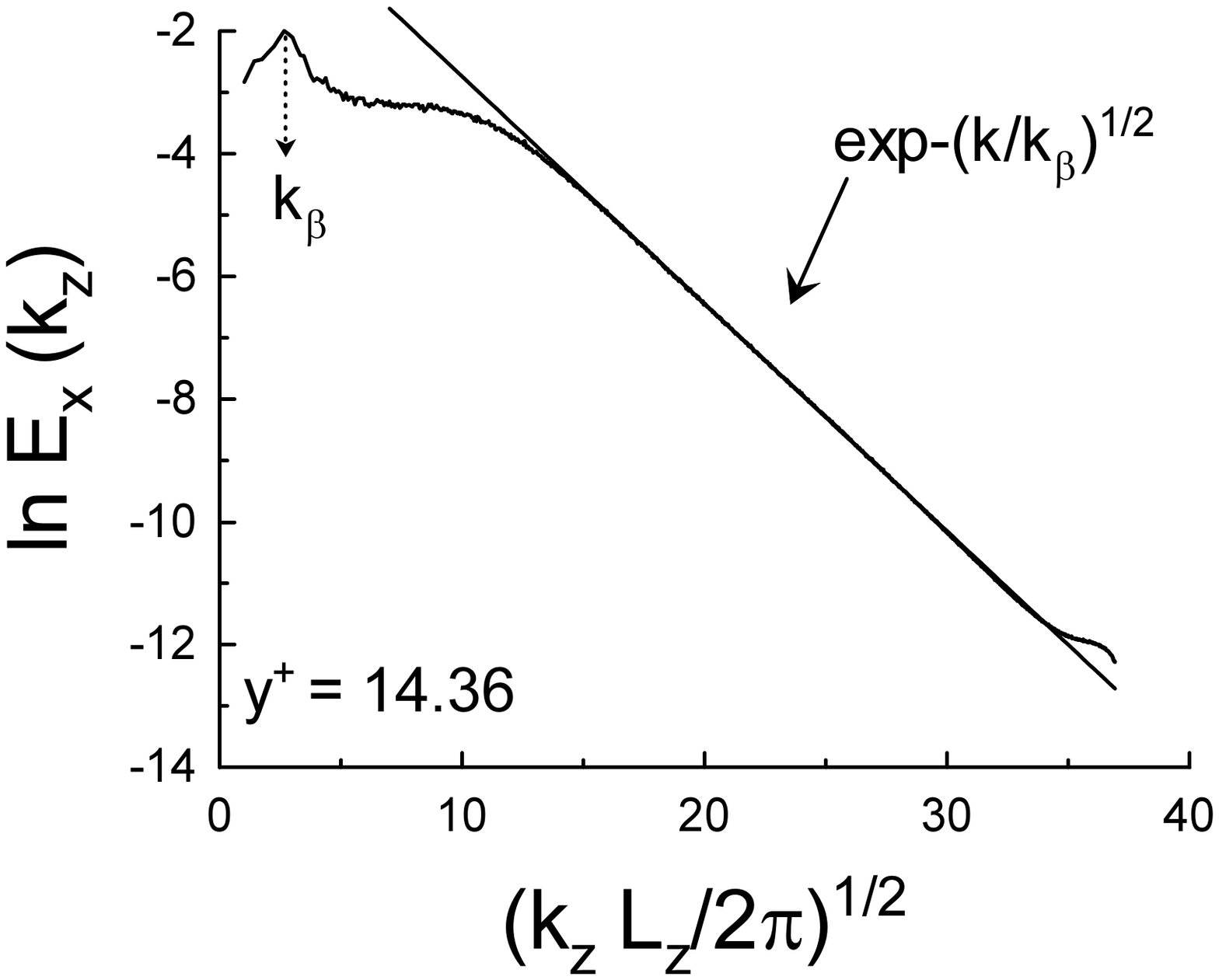}\vspace{-3cm}
\caption{\label{fig3} Logarithm of the power spectrum of streamwise component of velocity against of $(k_z L_z/2\pi)^{1/2}$ ($y^{+} =14.36$ and $L_z$ is the DNS box dimension). The data are from site \cite{tor}. The straight line corresponds to the exponential spectrum Eq. (5) with $\beta =1/2$.}
\end{center}
\end{figure}

\begin{figure}
\begin{center}
\includegraphics[width=8cm \vspace{-1.1cm}]{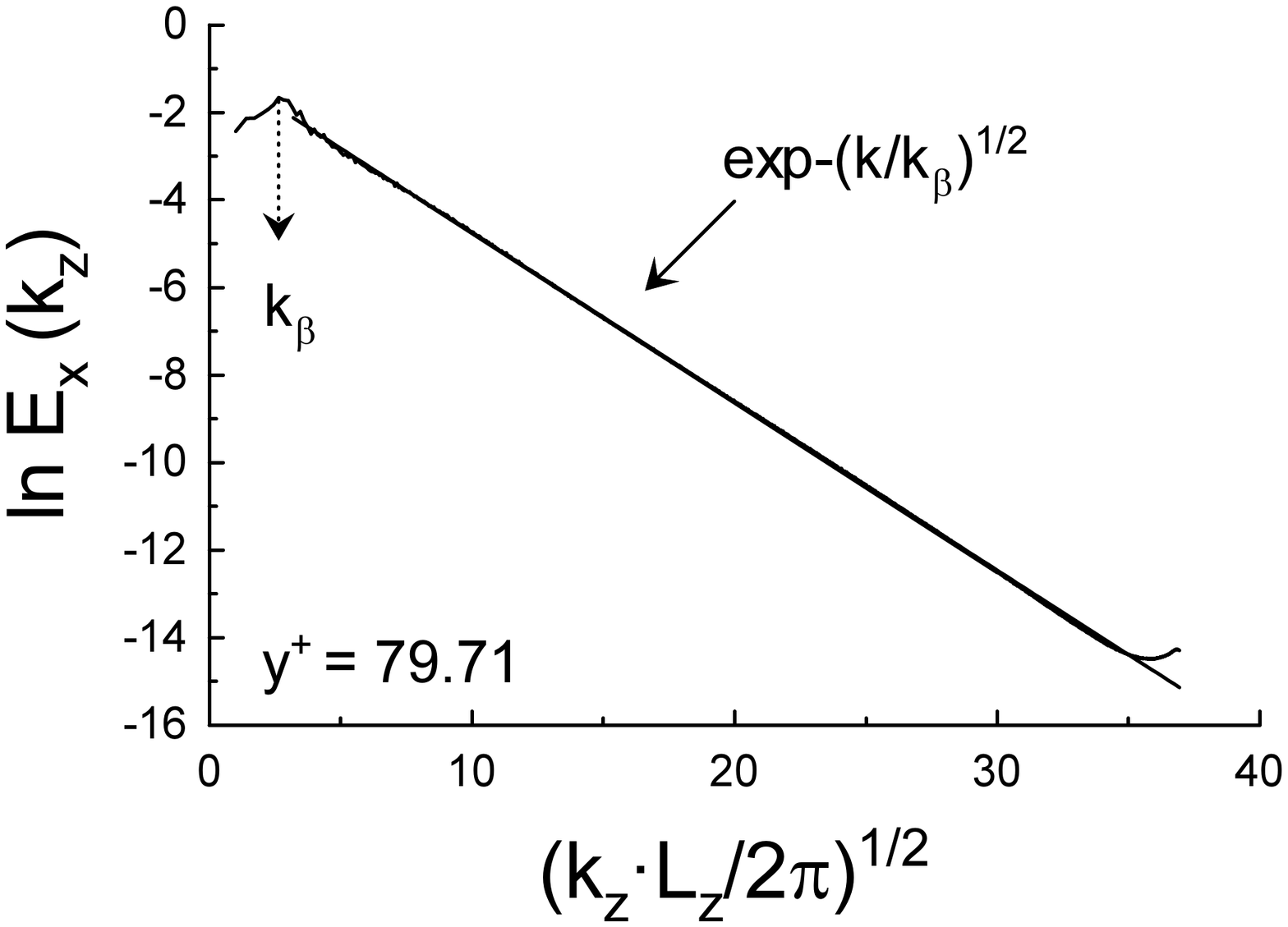}\vspace{-3.5cm}
\caption{\label{fig4} As in Fig. 3 but for $y^{+} =79.71$. } 
\end{center}
\end{figure}

   Figures 3-4 show the data of a direct numerical simulation (taken from the site \cite{tor}) of fully developed boundary layer flow at $Re_{\tau}\simeq 1990$ (see for more details Refs. \cite{sim}-\cite{sjm}). These spectra have a low wavenumber peak. Its position does not vary with the distance $y^{+}$ to the wall. In the range $14 < y^+ < 80$ (we will call this range as "range of coherence") the stretched exponential part of the spectra with $\beta =1/2$ (see Eq. (7)) exhibits tuning of the $k_{\beta}$ to the wavenumber equals to position of this spectral peak (see the arrows under the peak in Figs. 3 and 4). Figure 5 shows the streamwise  velocity {\it fluctuations} (in wall scaling and subtracting the maximum) for the data obtained in an experimental fully developed turbulent boundary layer \cite{gra}. The data were taken from Ref. \cite{sjm} and are in agreement with the DNS data (Fig. 8 in the Ref. \cite{sjm}). One can readily recognize the range of coherence. Outside of the range of coherence the tuning is broken.

\section{Turbulent thermal convection}
\begin{figure}
\begin{center}
\includegraphics[width=8cm \vspace{-1.05cm}]{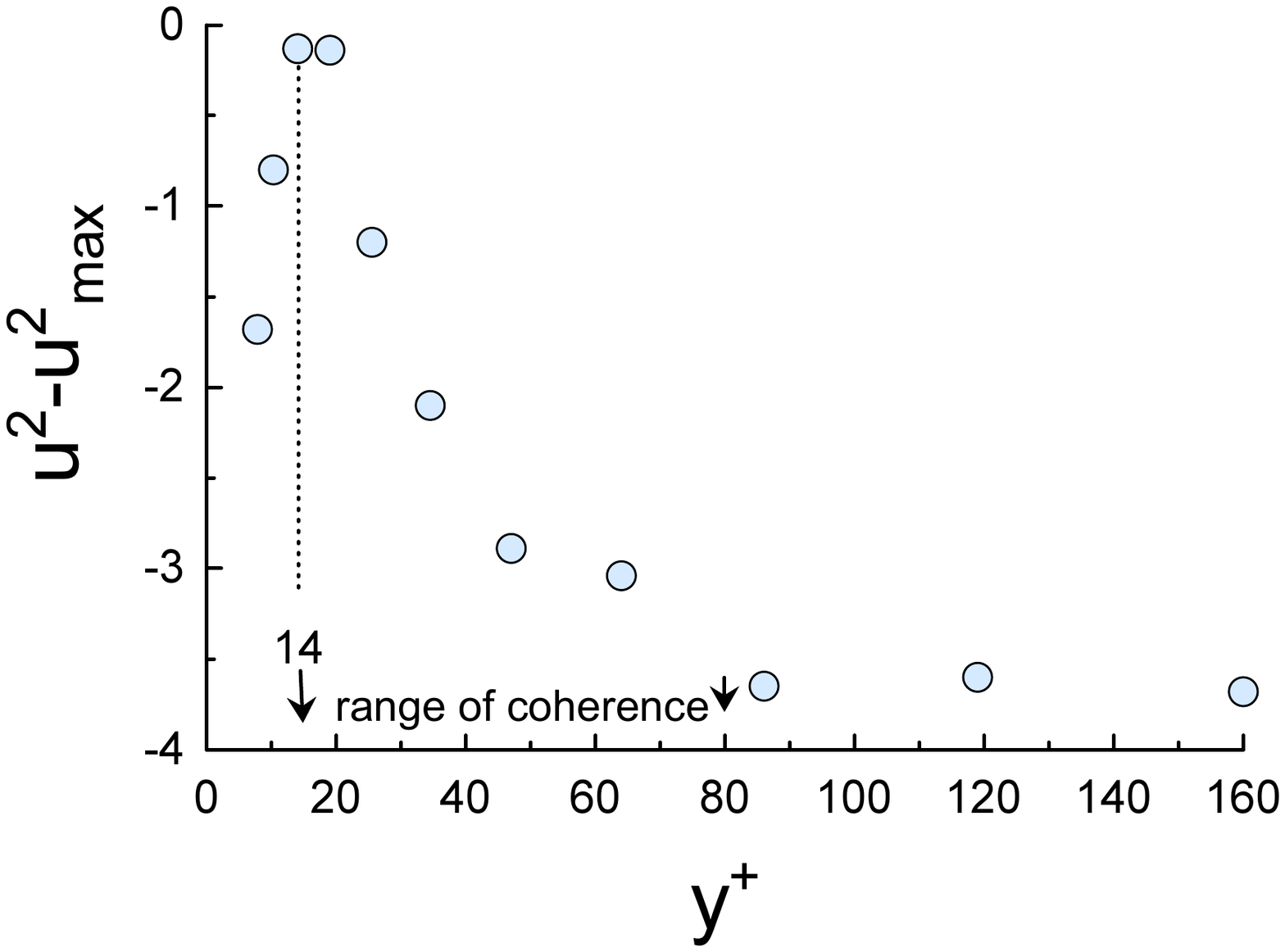}\vspace{-3.5cm}
\caption{\label{fig5} The range of coherence. The streamwise  velocity {\it fluctuations} (in wall scaling and subtracting the maximum) for the data obtained in an experimental fully developed turbulent boundary layer \cite{gra}. } 
\end{center}
\end{figure}

\begin{figure}
\begin{center}
\includegraphics[width=9cm \vspace{-2cm}]{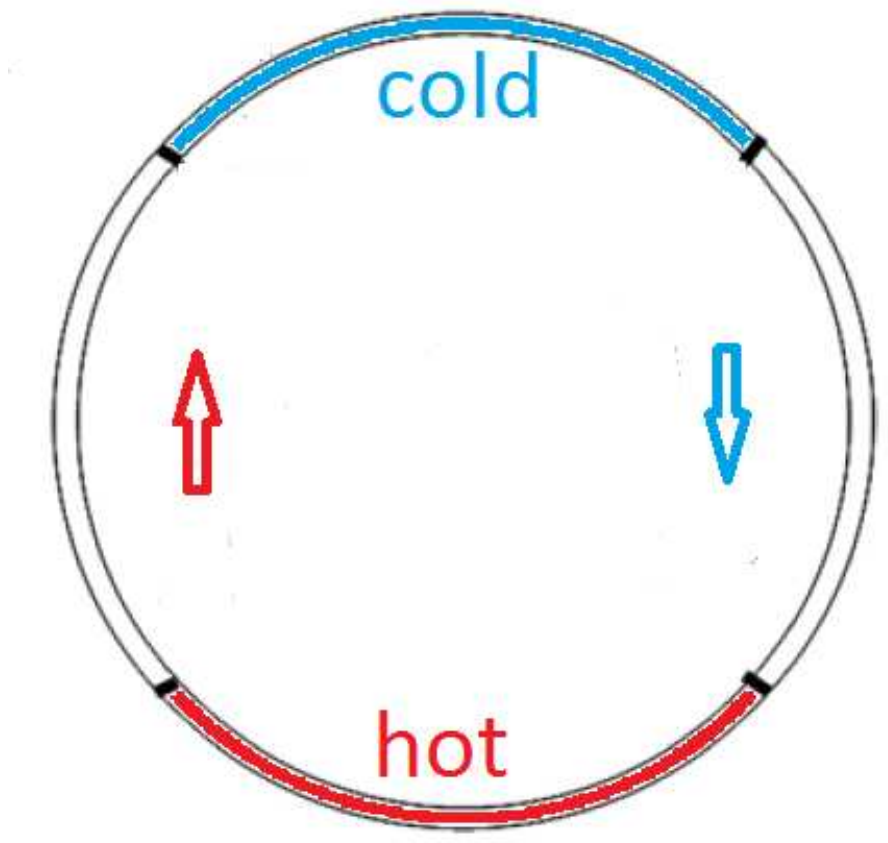}\vspace{-4.5cm}
\caption{\label{fig6} Sketch of the experimental situation presented in the Ref. \cite{song}.  } 
\end{center}
\end{figure}

 The Lorenz system discussed in the Introduction is a very simplified model for thermal convection.  A horizontal cylinder apparatus, sketched in Fig. 6,  was used recently for experimental study of real thermal convection \cite{song}. At high Rayleigh numbers a large scale circulation has been observed in the turbulent thermal convection (cf Ref. \cite{sbn} for upright cylinder apparatus). This circulation breaks the symmetry imposed by the ideal boundary conditions. The circulation is not stable and it changes its orientation with time (again a relexional symmetry is spontaneously broken by this process). As it is usual for the relexional symmetry breaking the large scale processes are statistically periodic (in order to restore the relexional symmetry statistically).  For parameters considered here both processes: the circulation and change of its orientation, have the same period corresponding to the position of the dominating peak ($f\simeq 0.08$ Hz) in Fig. 7, showing the horizontal velocity spectrum (the data are taken from the Ref. \cite{song}). The dashed line is drawn in Fig. 7 in order to indicate the stretched exponential spectrum Eq. (4) with $\beta =1/2$ (Eq. (7)). As for the Lorenz model the $f_{\beta}$ tuning to the large scale coherent motions (LSCM) has here subharmonic (period doubling) form: $f_{\beta} \simeq 0.04$ Hz. And the LSCM can be considered as a natural component of the distributed chaos in this case as well.  
\begin{figure}
\begin{center}
\includegraphics[width=8cm \vspace{-1.2cm}]{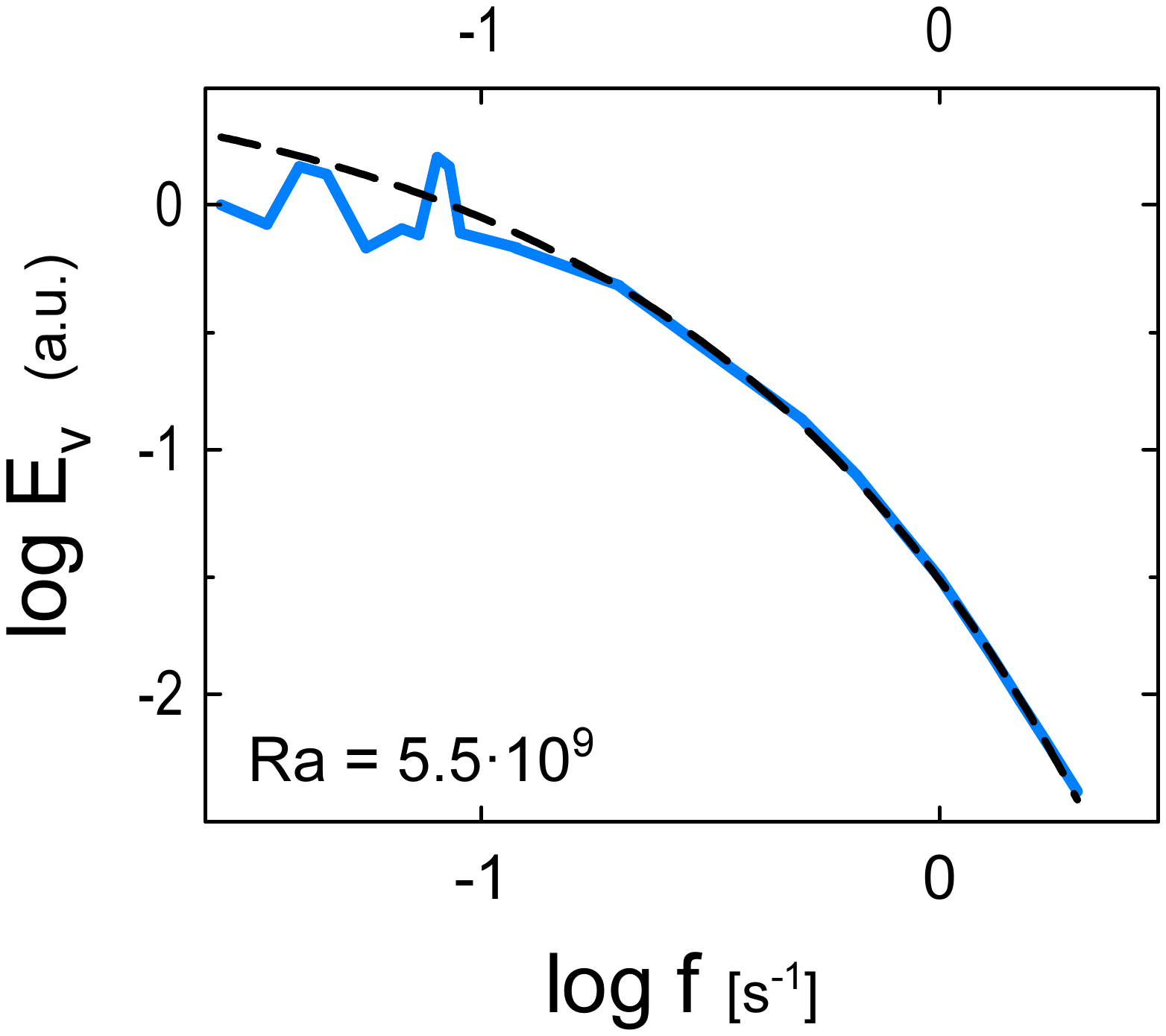}\vspace{-3.2cm}
\caption{\label{fig7} Logarithm of power spectrum of horizontal velocity against logarithm of frequency ($Ra=5.5\cdot 10^{9}$ and aspect ratio $\Gamma =0.5$). The data are taken from the Ref. \cite{song}). The dashed line is drawn to indicate the stretched exponential spectrum Eq. (4) with $\beta =1/2$ (Eq. (7)).} 
\end{center}
\end{figure}  

\section{Cuette-Taylor flow}  

The Cuette-Taylor flow between concentric cylinders (with rotating inner cylinder) was the first real hydrodynamic example of transition to turbulence via deterministic chaos with exponential broadband spectrum \cite{bs}. This situation corresponds to the probability density $P(f)$ in the Eq. (3) as a delta-function. With increasing Reynolds number the probability distribution $P(f)$ is widening. There is also a parallel process at the large scales, where the instabilities create and change coherent motions: vortexes and azimuthal (helical) waves. These two processes are closely interrelated because of fluxes and balance of momentum, angular momentum, energy and helicity \cite{l} along different scales. They reach a quasi-equilibrium controlled by the Levich-Tsinober (helicity) integral Eq. (8) with $f_{\beta}$ tuned to the characteristic frequency of the azimuthal (helical) waves. Since the azimuthal waves (LSCM in this case) are a natural component of the distributed chaos here one can expect that the tuning will be subharmonic (period doubling).

\begin{figure}
\begin{center}
\includegraphics[width=8cm \vspace{-1.2cm}]{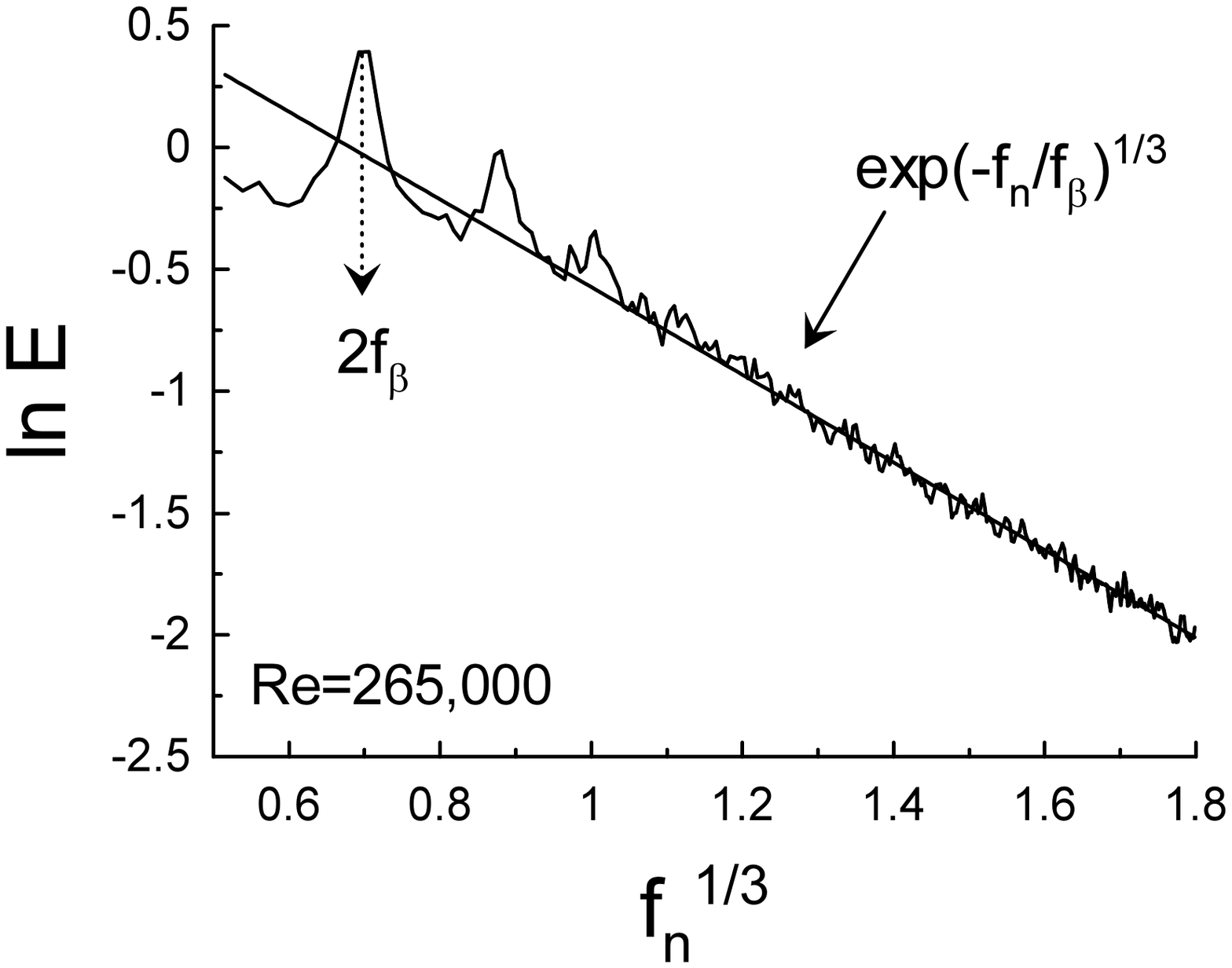}\vspace{-3.5cm}
\caption{\label{fig8} Logarithm of spectrum of the azimuthal component of velocity against $f_n^{1/3}$ ($f_n=2\pi f/\Omega$, where $\Omega$ is angular rotation rate of inner cylinder). The subharmonical tuning has been indicated by an arrow under the fundamental harmonic of the azimuthal waves.  } 
\end{center}
\end{figure}
\begin{figure}
\begin{center}
\includegraphics[width=8cm \vspace{-1.4cm}]{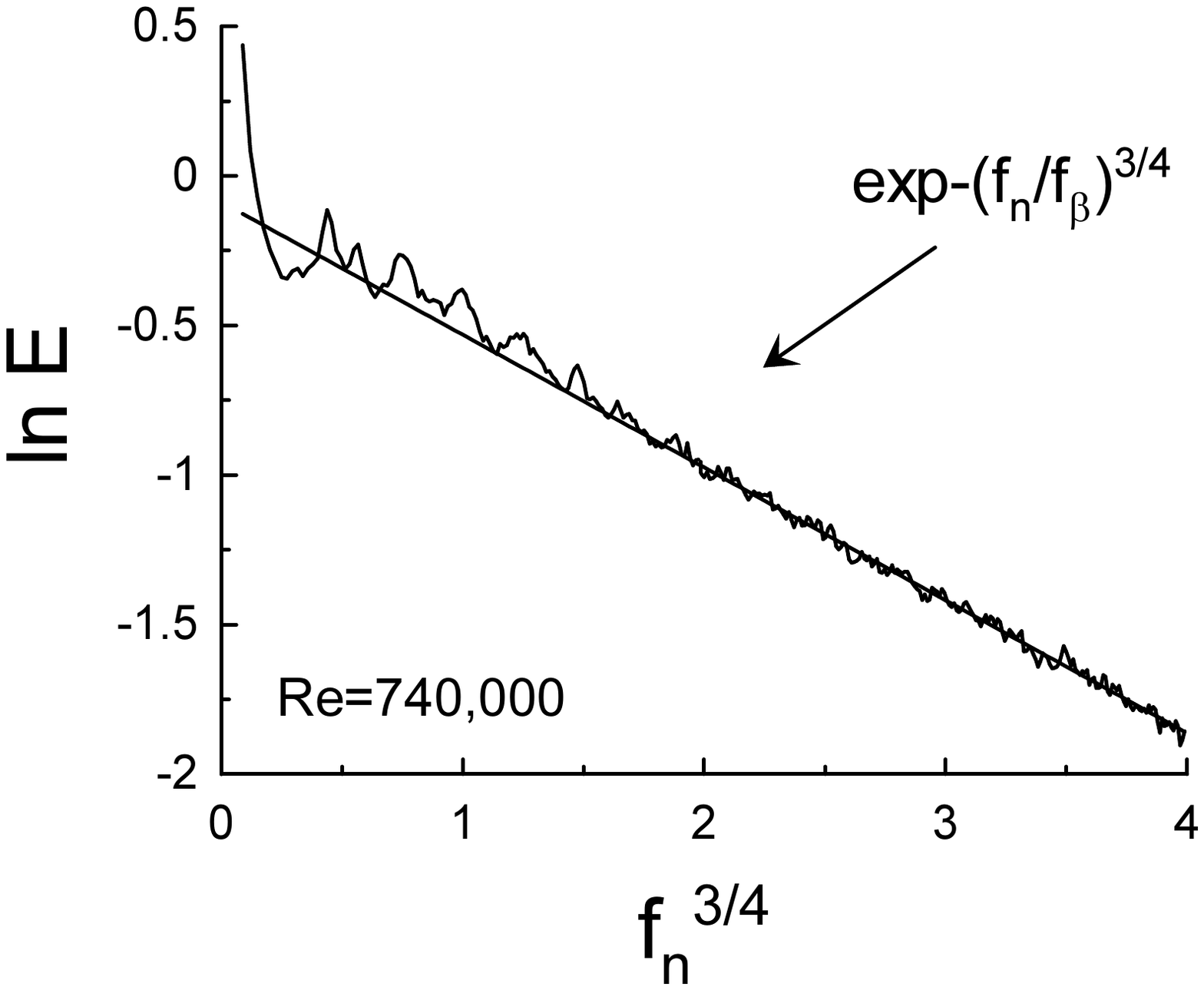}\vspace{-3.2cm}
\caption{\label{fig9} As in Fig. 8 but for $Re=740,000$. Note the change in value of $\beta$.} 
\end{center}
\end{figure}  
  Figures 8 shows power spectrum of azimuthal component of velocity for an experimental Cuette-Taylor flow at Reynolds number $Re=265,000$ (the data are taken from Ref. \cite{ls}, fig. 6).  The straight line corresponds to the stretched exponential Eq. (4) with $\beta = 1/3$ (Eq. (8)). The peaks in this figure correspond to fundamental frequency of the large scale azimuthal waves and its harmonics. The subharmonic (period-doubling) tuning of the $f_{\beta}$ has been also indicated in the Fig. 8 by an arrow under the fundamental frequency peak of the azimuthal waves. 
  
  With further increase of the Reynolds number another equilibrium (local homogeneity and isotropy) is reached. The straight line in Fig. 9 ($Re=740,000$) corresponds to the stretched exponential Eq. (4) with $\beta = 3/4$ (Eq. (6)).  No inertial range is developing here. The large scales (low frequencies) are still dominated by the azimuthal waves, but the tuning is gone, due to the local (small-scale \cite{my}) isotropy and homogeneity at the large Reynolds number.

\section{Acknowledgement}

I thank G. Borrell, J. Jimenez, S. Hoyas, G. S. Lewis, R. D. Moser,  Y. Mizuno, J. A. Sillero, and  M. P. Simens  for sharing their data.

\end{document}